\documentclass[a4paper]{article}
\usepackage{graphicx}
\usepackage{amsmath}
\usepackage{bm}
\usepackage[hypertex]{hyperref}

\newcommand{\bs}[1]{\boldsymbol{#1}}
\newcommand{\DR}{D_{\rm R}}

\author{J. T. Locsei and T. J. Pedley}
\title{Bacterial tracking of motile algae assisted by algal cell's vorticity field}
\begin{document}
\maketitle

A short version of the title for running headings is `Bacterial
tracking of motile algae'.
\\
\\
Both authors are based at the \vspace{0.5 cm} \\%
Department of Applied Mathematics and Theoretical Physics \\*%
Centre for Mathematical Sciences \\*%
University of Cambridge \\*%
Cambridge CB3 0WA \\*%
United Kingdom
\\
\\
J. T. Locsei is the corresponding author, and may be contacted at
j.t.locsei@damtp.cam.ac.uk or +44 (0)7849829002

\newpage

\begin{abstract}

Previously published experimental work by other authors has shown that
certain motile marine bacteria are able to track free swimming algae by
executing a zigzag path and steering toward the algae at each turn.
Here, we propose that the apparent steering behaviour could be a
hydrodynamic effect, whereby an algal cell's vorticity and strain-rate
fields rotate a pursuing bacterial cell in the appropriate direction.
Using simplified models for the bacterial and algal cells, we
numerically compute the trajectory of a bacterial cell and demonstrate
the plausibility of this hypothesis.

\end{abstract}

\newpage

\section{Introduction}

The experiments of Barbara and Mitchell \cite{Barbara.2003} demonstrate
that certain species of marine bacteria are able to track motile algae
by a combination of reversing direction and steering, presumably to
make use of nutrients exuded by the algae. Their results raise a number
of interesting questions about how the bacteria are able to perform
this feat, such as (i) how do they steer, and (ii) how do they decide
when to reverse. In this paper we shall address the first of these
questions: steering. We hypothesise that a bacterial cell exploits the
vorticity and strain-rate fields generated by an algal cell in order to
steer. Before presenting our calculations, we give a brief overview of
bacterial chemotaxis.

Bacteria play an important role in marine ecosystems. Field studies
indicate that approximately half of oceanic primary production (carbon
fixed by phytoplankton) is channelled via bacteria into the microbial
loop of the pelagic food web (\cite{Azam.1998} and references therein).
This remarkable rate of bacteria-mediated transformation of organic
matter is facilitated by bacterial motility. It is estimated that 20 to
70\% of planktonic bacteria are motile \cite{Grossart.2001,
Fenchel.2001}. Motility allows marine bacteria to locate sources of
nutrients, for instance by colonising falling sinking aggregates
(`marine snow') \cite{Kiorboe.2002}, or tracking motile marine algae
\cite{Barbara.2003} which leak organic matter \cite{Mague.1980}.

The most studied and well-understood motile bacterium is
\emph{Escherichia coli}, an enteric bacterium, but marine bacteria
differ from \emph{E. coli} in several important ways. First, many
species of marine bacteria swim much faster than \emph{E. coli}, with
maximum speeds up to $400 \,{\rm \mu m\,s}^{-1}$ \cite{Magariyama.1994,
Mitchell.1995b}, compared to typical speeds of $30\,{\rm \mu
m\,s}^{-1}$ for \emph{E. coli} \cite{Darnton.2007}. Second, marine
bacteria are able to respond to changes in chemoattractant
concentrations within a fraction of a second, as evidenced by their
ability to form tight clusters \cite{Mitchell.1996}, whereas \emph{E
coli} respond to changes in chemoattractant concentrations over the
course of several seconds \cite{Segall.1986}. Finally, whereas \emph{E.
coli} moves in a run-and-tumble motion \cite{Berg.1972}, many strains
of marine bacteria move in a run-and-reverse motion
\cite{Mitchell.1991, Mitchell.1996}. Mechanistically, this is explained
by the fact that \emph{E. coli} has several flagella which may either
bundle together to propel the cell in a straight `run', or fly apart to
make the cell `tumble' \cite{Turner.2000}, whereas many marine bacteria
have only a single polar flagellum which may spin either one way or the
other for a `run' or `reverse'. (For a review of other types of
bacterial motions, see \cite{Mitchell.2006}.) It has been suggested
that run-and-reverse motion favoured by marine bacteria allows them to
perform chemotaxis (swim toward a food source) more effectively in a
turbulent environment, compared to run-and-tumble motion
\cite{Luchsinger.1999}. Other theoretical studies indicate that
run-and-reverse may be a favourable strategy when moving up a uniform
chemoattractant gradient in a shear flow \cite{Locsei.2008} or when an
organism is capable of direct gradient detection \cite{Locsei.2007}.

\begin{figure}
   \centerline{
      \includegraphics[width=0.7\linewidth,height=!]{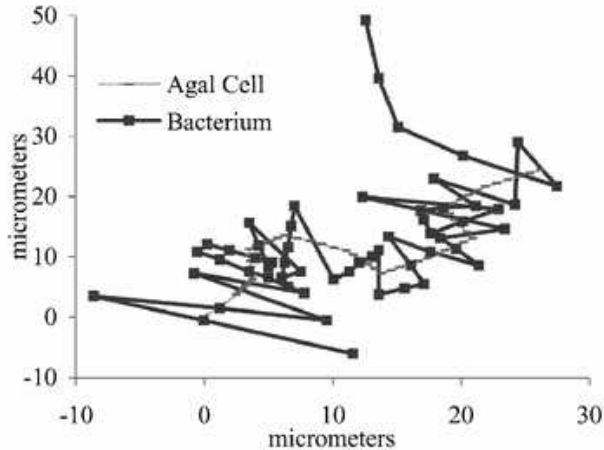}
   }
   \caption{
      Observations of the paths of a \emph{P. lutheri} algal cell being
      tracked by a \emph{P. haloplanktis} bacterial cell (reproduced
      from figure 1a of \cite{Barbara.2003}, with permission). The
      algal cell track starts at the bottom-left of the figure. Each
      square represents a single data point; data was collected at $24$
      frames/sec.
   }
   \label{fig:fig1a_from_Barbara}
\end{figure}

It seems that the run-and-reverse motion of marine bacteria is also
advantageous when following a moving target such as a motile algal
cell. Barbara and Mitchell \cite{Barbara.2003} reported that motile
marine bacteria of the species \emph{Pseudoalteromonas haloplanktis}
and \emph{Shewanella putrefaciens} were able to track motile marine
algae of the species \emph{Pavlova lutheri} by moving in zigzag paths
and apparently steering, consecutively turning up to 12 times toward
the algae (figure \ref{fig:fig1a_from_Barbara}). However, \emph{P.
haloplanktis} and \emph{S. putrefaciens} cells have a single polar
flagellum \cite{Gauthier.1995, Holt.2005}, and it is not clear how they
could actively steer with such an apparatus. Here, we suggest that the
apparent steering is in fact a passive hydrodynamic effect, whereby an
algal cell's vorticity and strain rate fields rotate the pursuing
bacterial cell in the appropriate direction. Sections
\ref{subsec:algalcell} and \ref{subsec:modelbac} introduce simplified
models for the bacterial and algal cells, based on what is known about
them from experiments, and section \ref{subsec:scaling} presents a
scaling argument to show that the algal cell's vorticity field has the
right order of magnitude to account for the observed bacterial
steering. Section \ref{subsec:simulations} describes a method for
simulating bacterial tracking, and section \ref{sec:results} presents
simulation results and investigates how parameters such as swimming
speeds influence the success of tracking. Section \ref{sec:discussion}
summarises the main findings and their biological relevance.

\section{Methods}
\label{sec:methods}

\subsection{Model algal cell}
\label{subsec:algalcell}

The \emph{P. lutheri} algal cells used in the experiments of
\cite{Barbara.2003} had diameters of approximately $6\;{\rm \mu m}$ and
swam at average speeds of $v_{\rm alg} \approx 40\, {\rm \mu m\,
s}^{-1}$. They changed direction about three times per second, and
bacterial cells were able to track them despite these turns, but we
ignore this complication and let the model algal cell swim in a
straight line. All algae of the order Pavlovales have approximately
spherical cell bodies and two flagella of unequal length, the longer of
which is covered in fine hairs \cite{Goldstein.1992}. The flagella
lengths of \emph{P. lutheri} have not been measured, but in the related
species \emph{P. gyrans} the length of the longer flagellum is about
$10$--$20\,{\rm \mu m}$, and in \emph{P. mesolychnon} the length of the
longer flagellum is about $15$--$20\,{\rm \mu m}$. The details of the
flagellar beat pattern of \emph{P. lutheri} are not well characterised,
but it is known that the longer flagellum is directed forward during
swimming. Thus, a \emph{P. lutheri} cell is a `puller', since it pulls
itself forward with its longer flagellum, as opposed to a `pusher' such
as \emph{Escherichia coli} whose propulsive flagella bundle lies behind
the cell as it swims.

\begin{figure}
   \centerline{
      \includegraphics[width=0.5\linewidth,height=!]{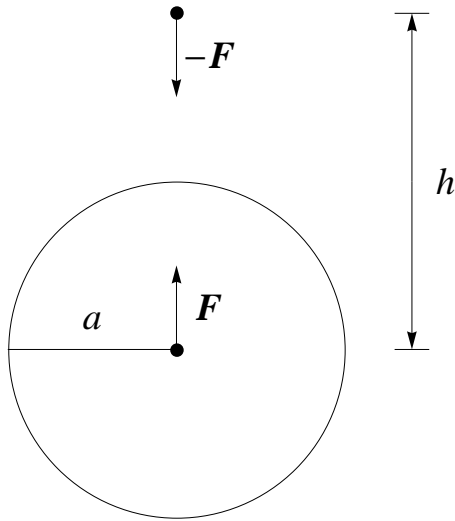}
   }
   \caption{
      Simplified model of a \emph{P. lutheri} algal cell, consisting of
      a sphere of radius $a$ acted upon by a force $\bs{F}$ and a
      Stokeslet of strength $-\bs{F}$ located a distance $h$ ahead of
      the sphere (the `phantom flagellum'), such that the cell as a
      whole is force-free. In the orientation depicted here, the cell
      is swimming upward.
   }
   \label{fig:algae_model}
\end{figure}

In the absence of detailed knowledge of the flagellar beat pattern, we
use a simplified model for a \emph{P. lutheri} cell (figure
\ref{fig:algae_model}). The cell body is treated as a sphere of radius
$a=3\,{\rm \mu m}$ acted upon by a force $\bs{F}$ in the swimming
direction - this force is intended to model the tension at the base of
the flagellum. The influence of the flagellum on the fluid is modelled
as a point force $\bs{-F}$ acting on the fluid at a distance $h$ ahead
of the cell body, with $h-a$ equal to approximately half the length of
the longer flagellum. (This point force `phantom' flagellum is
analogous to that used in \cite{Hernandez-Ortiz.2005} for modelling
motile bacteria.) We choose $h = 10\,{\rm \mu m}$, but our results are
qualitatively unchanged for $h$ as low as $\rm 1\,{\rm \mu m}$ and as
high as $20\,{\rm \mu m}$. Note the algal cell's velocity field in this
model is constant in time, whereas in reality it is likely to vary
periodically in time with the flagellar beat pattern, as well as vary
over a longer time-scale as the algal cell changes its swimming speed.

\begin{figure}
   \centerline{
      \includegraphics[width=0.5\linewidth,height=!]{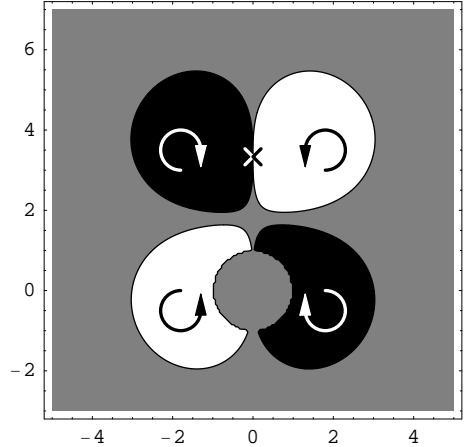}
   }
   \caption{
      Cross section of the vorticity field around the model algal cell.
      Black and white regions are where the magnitude of the vorticity
      field exceeds $0.2 v_{\rm alg} / a$, and grey regions are where
      the magnitude of the vorticity field is less than $0.2\, v_{\rm
      alg} / a$. Arrows show the vorticity direction. The cell body is
      centered on the origin, and the phantom flagellum is marked by a
      cross. Scale is in units of the cell radius $a$.
   }
   \label{fig:vorticitylobes}
\end{figure}

\begin{figure}
   \centerline{
      \includegraphics[width=0.5\linewidth,height=!]{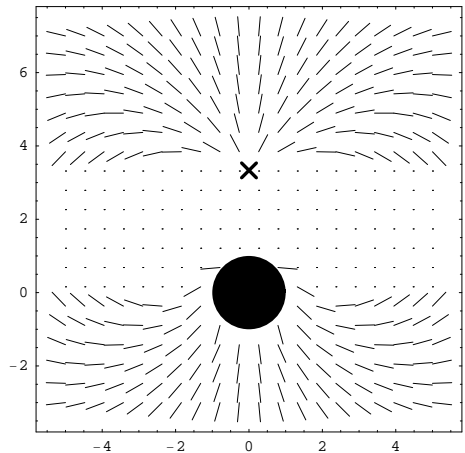}
   }
   \caption{
      Direction of maximum extension rate of the strain rate field of
      the model algal cell. (An elongated body, such as a pursuing
      bacterium, will tend to align with the direction of maximum
      extension rate.) The cell body is shown as a black disk and the
      phantom flagellum is marked by a cross. Scale is in units of the
      cell radius $a$. In the blank regions, the direction of maximum
      extension rate is perpendicular to the page.
   }
   \label{fig:strainratefield}
\end{figure}

The Reynolds number for the algal cell, based on the cell diameter,
swimming speed and the viscosity and density of water, is ${\rm Re}
\approx 10^{-4}$. Hence, the fluid flow around the cell is viscous,
non-inertial, and governed by the Stokes equations, assuming that the
fluid is Newtonian. Using the linearity of the Stokes equations, the
velocity field around the algal cell is obtained by adding the solution
for a translating sphere to the solution for a Stokeslet outside a
stationary sphere \cite{Kim.1992}; the resulting expression is lengthy
and is omitted here. Figure \ref{fig:vorticitylobes} shows the
vorticity field around the model algal cell, and figure
\ref{fig:strainratefield} shows the direction of maximum extension rate
of the strain rate field of the model algal cell.

\subsection{Model bacterial cell}

\label{subsec:modelbac}

The two species of bacteria used in the experiments of
\cite{Barbara.2003}, \emph{Pseudoalteromonas haloplanktis} and
\emph{Shewanella putrefaciens}, swam at average speeds of $v_{\rm bac}
\approx 200 \; {\rm \mu m\,s}^{-1}$ while tracking algae. Cells from
each species have a single, helical, polar flagellum
\cite{Gauthier.1995, Holt.2005}, which may rotate either
counterclockwise or clockwise for a `run' or `reverse'. Barbara and
Mitchell \cite{Barbara.2003} did not report the size of the bacterial
cells used in their study, but cells of all \emph{Pseudoalteromonas}
species are rod shaped with width 0.2 to 1.5 ${\rm \mu m}$ and length
1.8 to 3 ${\rm \mu m}$ \cite{Gauthier.1995}, while those of the
\emph{Shewanella} species are rod shaped with width 0.4 to 0.7 ${\rm
\mu m}$ and length 2 to 3 ${\rm \mu m}$ \cite{Venkateswaran.1999}. In
each case the length of the flagellum is a few times that of the cell
body. For the purpose of modelling we treat a swimming bacterial cell
as a prolate spheroid of infinitesimal size which translates along its
long axis and is rotated and advected by the velocity  field of the
algal cell. Ideally, one should incorporate the finite size of the
bacterial cell and the influence of its velocity field on the algal
cell, but to simplify the mathematics we neglect these complications.

Cells of \emph{E. coli}, which are a similar size to \emph{P.
Haloplanktis} and \emph{S. putrefaciens}, exhibit small, random changes
in direction, and these have been attributed to rotational Brownian
motion (i.e. thermal collisions with molecules in the surrounding
fluid). Berg \cite{Berg.1983} estimated the the rotational diffusion
coefficient due to Brownian rotation to be $\DR \approx 0.062\;{\rm
radians}^2{\rm s}^{-1}$ for an \emph{E. coli} cell swimming in a fluid
of viscosity 2.7 cp at $32^{\rm o}{\rm C}$, and this is consistent with
experimental observations \cite{Berg.1972}. However, this estimate was
based on treating the \emph{E. coli} cell as a sphere of diameter
$2\;{\rm \mu m}$, and when one takes into account the stabilising
effect of the flagellar bundle, the theoretically predicted rotational
diffusion coefficient is an order of magnitude smaller than the
observed rotational diffusivity (H. Fu, personal communication), so it
seems likely that the observed rotational diffusivity is in fact due to
intrinsic `wobbly swimming' rather than true (thermal) Brownian
rotation (H. Berg, personal communication). It is therefore unclear
what value of $\DR$ is appropriate for \emph{P. Haloplanktis} and
\emph{S. putrefaciens}. In section \ref{sec:results} we shall simply
use a default value of $\DR = 0.062\;{\rm radians}^2{\rm s}^{-1}$, and
demonstrate that the results have only a weak dependence on $\DR$
anyway.


Given our assumptions, the translational velocity of the model
bacterial cell is
\begin{equation}
\label{eq:xdot}
   \dot{\bs{x}}= v_{\rm bac} \bs{p} + \bs{u}(\bs{x},t),
\end{equation}
where $\bs{x}$ is the position vector of the bacterial cell, $v_{\rm
bac}$ is the bacterial cell's swimming speed, $\bs{p}$ is a unit vector
denoting the swimming direction of the bacterial cell, and
$\bs{u}(\bs{x},t)$ is the velocity field due to the algal cell at the
position of the algal cell. We treat $v_{\rm bac}$ as a constant,
although in the experiments of \cite{Barbara.2003} the bacterial cells
often altered their speeds by a factor of $2$ or more during tracking.

With the exception of reversals and collisions (discussed later), the
rate of change of direction of the bacterial cell is
\begin{equation}
\label{eq:pdot}
   \dot{\bs{p}} = \frac{1}{2} \bs{\omega} \times \bs{p} + \alpha_0\,
   \bs{p} \cdot \bs{E} \cdot (\bs{I}-\bs{p}\bs{p})
   + \bs{\xi} \times \bs{p}.
\end{equation}
The first two summands on the right hand side of (\ref{eq:pdot})
represent the deterministic rotation of the bacterial cell by the
vorticity field $\bs{\omega}$ and strain-rate field $\bs{E}$ of the
algal cell \cite{Leal.1972}, and
\begin{equation}
\label{eq:alpha}
   \alpha_0 = (\eta^2-1)/(\eta^2+1),
\end{equation}
where $\eta$ is the ratio of the major to minor axis of the cell
(slenderness ratio). We choose $\eta = 10$ as a default value
(suggested as a reasonable value by \cite{Locsei.2008}), and we explore
the effect of different values for $\eta$ in section \ref{sec:results}.
The last term on the right hand size of (\ref{eq:pdot}) represents
rotational diffusivity, with $\bs{\xi}$ being a random, white-noise
angular velocity with autocorrelation function
\begin{equation}
   \langle \xi_i(t_1) \xi_j(t_2) \rangle
      = \DR \delta_{ij} \delta(t_1-t_2).
\end{equation}

An important ingredient of the model is a rule to decide when the
bacterial cell reverses direction. The details of the chemotactic
response of marine bacteria are yet to be characterised
\cite{Mitchell.1996}. For instance, it is not known whether they
perform temporal comparisons of chemoattractant concentrations like
\emph{E. coli} \cite{Segall.1986}, whether they directly detect
chemoattractant gradients, or whether they respond to absolute
chemoattractant concentrations. Barbara and Mitchell
\cite{Barbara.2003} observed that during tracking a bacterial cell
remained within a distance of about $6 \,{\rm \mu m}$ from the algal
cell. Based on this, we choose the simplest rule for bacterial
reversals, and make the model bacterial cell reverse direction whenever
it reaches a `reversal distance' $R \approx 6\,{\rm \mu m}$ from the
centre of the model algal cell.

One way that the bacterial cell might achieve this in reality is to
reverse direction whenever the chemoattractant concentration drops
below a certain threshold. The chemoattractant in the experiments is
likely to have been dissolved oxygen released by the algal cells (J. G.
Mitchell, personal communication), which has a diffusion coefficient of
approximately $2000\, {\rm \mu m^2 \, s^{-1}}$ in water at room
temperature. Given an algal cell swimming speed of $40\,{\rm \mu m\,
s^{-1}}$ and an algal cell radius of $3\,{\rm \mu m}$, the P\'{e}clet
number is ${\rm Pe} \approx 0.06 \ll 1$. If the chemoattractant is a
small organic molecule instead, then a commonly used diffusion
coefficient is $1000\, {\rm \mu m^2 \, s^{-1}}$ (used for instance in
\cite{Purcell.1977, Jackson.1987a, Jackson.1989, Luchsinger.1999,
Bowen.1993}), which gives ${\rm Pe} \approx 0.12 \ll 1$. In either
case, diffusion dominates over advection and to leading order the
distribution of chemoattractant concentration $c$ in the near vicinity
of the algal cell is spherically symmetric with distribution
\begin{equation}
   c(r) = c_0 a/r,
\end{equation}
where $c_0$ is the concentration on the surface of the algal cell and
$r$ is distance from the centre of the algal cell. So, if the bacterial
cell reverses direction when, say, $r=R=2a$, the corresponding
concentration threshold is $c_0/2$.

The final ingredient for the model bacterial cell is a rule for what
happens when it collides with the algal cell. In reality, there will be
some complicated interaction involving a lubrication flow between the
two cells, and potentially also an active response from one or both of
the cells. Hydrodynamic effects are know to cause bacteria propelled by
helical flagella to swim along curved paths when they are near a planar
surface (\cite{DiLuzio.2005} and references therein), and a similar
effect might occur in the interaction between the bacterial and algal
cells. For the purpose of the model, however, we assume that when the
bacterial cell collides with the algal cell, it simply glances off and
its swimming direction changes so as to be parallel with the surface of
the algal cell. Mathematically, the bacterial swimming direction
$\bs{p}_1$ after collision is related to the swimming direction
$\bs{p}_0$ before collision by
\begin{equation}
   \bs{p}_1 = \frac{\bs{u}}{\sqrt{\bs{u} \cdot \bs{u}}},
\end{equation}
where
\begin{equation}
   \bs{u} = (\bs{I}-\bs{n}\bs{n})\cdot\bs{p}_0,
\end{equation}
$\bs{I}$ is the identity matrix, and $\bs{n}$ is the outward unit
normal vector from the algal cell surface.

\subsection{Scaling argument} \label{subsec:scaling}

\begin{figure}
   \centerline{
      \includegraphics[width=0.5\linewidth,height=!]{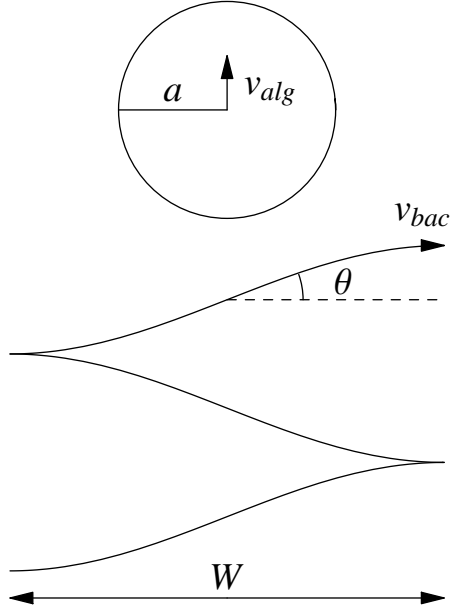}
   }
   \caption{
      Idealised zigzag path of bacteria tracking algae.
   }
   \label{fig:zigzagfig}
\end{figure}

Imagine a bacterial cell performing run-and-reverse motion in a side to
side fashion (with respect to the algal cell's velocity) just behind a
swimming algal cell. The algal cell's vorticity field (figure
\ref{fig:vorticitylobes}) rotates the bacterial cell such that instead
of retracing the same path over and over, the bacterial cell traces out
a curved zigzag path as depicted in figure \ref{fig:zigzagfig}, with a
net motion in the same direction as the algal cell's velocity.

A simple scaling argument suggests that the algal cell's vorticity
field alters the bacterial cell's path by the right order of magnitude
for the bacterial cell to keep pace with the algal cell. If the
bacterial cell is within a distance $R$ of the algal cell, then the
magnitude of the vorticity field it experiences is $\omega \sim v_{\rm
alg} / R$, where `$\sim$' means `is of order'. Let $W$ be the width of
the zigzag path followed by the bacterial cell, and let $\theta$ be the
angle between the bacteria cell's swimming direction and the plane
normal to the algal cell's swimming direction (see figure
\ref{fig:zigzagfig}). The angle $\theta$ is approximately the angle
through which the bacterial cell is rotated by the vorticity field as
it swims the width of the zigzag, so $\theta \sim \omega W / v_{\rm
bac} \sim (v_{\rm alg}/v_{\rm bac})W/R \sim v_{\rm alg}/v_{\rm bac}$,
since $W \sim R$. Let us assume that $v_{\rm bac} \gg v_{\rm alg}$, so
that $\theta \ll 1$. Then, the component of the bacterial cell's
velocity parallel to the algal cell's velocity is $v_{\rm bac}
\sin\theta \sim v_{\rm bac}\theta \sim v_{\rm alg}$. Hence the algal
cell's vorticity field has the right order of magnitude for the
bacterial cell to keep pace with the algal cell. Inserting numbers,
$v_{\rm alg} \approx 40\, {\rm \mu m\, s}^{-1}$ and $v_{\rm bac}
\approx 200\, {\rm \mu m\, s}^{-1}$, so $\theta \approx 40/200 = 0.2
\;{\rm radians} \approx 11^{\rm o}$. While this scaling argument
focusses on the effect of vorticity, the simulation results in section
\ref{sec:results} show that advection by the algal cell's velocity
field and rotation by its strain rate field are also important.

\subsection{Simulation method}

\label{subsec:simulations}

It is common practice to non-dimensionalise the quantities in a problem
before performing simulations. In the problem considered here, there
are six dimensional quantities of interest (five parameters plus time):
algal cell radius $a$, algal cell swimming speed $v_{\rm alg}$,
bacterial cell swimming speed $v_{\rm bac}$, reversal distance $R$,
bacterial cell rotational diffusivity $\DR$, and time $t$. In
principle, by non-dimensionalising with respect to, say, the algal cell
radius and the algal cell speed, one could reduce the number of
quantities to just four. In practice, however, this is impractical,
because then it is difficult to ascertain the effect of altering a
single dimensional parameter. For this reason we shall define a
non-dimensional time $t^* = t v_{\rm alg}/a$, but leave other
quantities dimensional. One unit of $t^*$ then corresponds to the time
it takes the algal cell to swim a distance equal to its own radius.

\begin{figure}
   \centerline{
      \includegraphics[width=0.5\linewidth,height=!]{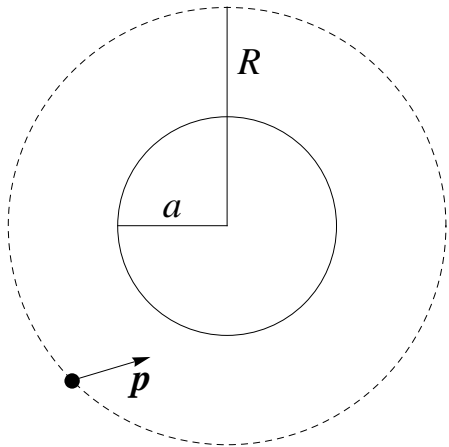}
   }
   \caption{
      Example initial condition. Circle of radius $a$ denotes the algal
      cell. Bacterial cell initial position is random and lies a distance $R$ from
      the centre of the algal cell. Bacterial initial swimming direction $\bs{p}$ is
      random and points inward with respect to the surface $r=R$ (dashed circle).
   }
   \label{fig:initial_condition}
\end{figure}

For each simulation, the initial position of the model bacterial cell
is chosen from a uniform, random distribution on the surface $r=R$,
where $r$ denotes distance from the centre of the algal cell. The
initial swimming direction of the bacterial cell is chosen from a
uniform, random distribution on the hemisphere of unit vectors that
point inward with respect to the surface $r=R$. An example initial
condition is shown in figure \ref{fig:initial_condition}. The algal
cell position and the bacterial cell position and orientation are then
evolved in time, using a simple Euler time-step scheme, according to
the rules outlined in section \ref{subsec:modelbac}. The time-step
$\Delta t$ is chosen to be $\Delta t = 0.005 a / v_{\rm alg}$.
(Simulations were repeated using smaller time-steps to check
convergence of results.) Brownian rotation is simulated by giving the
bacterial cell a new swimming direction at each time-step, chosen from
an axisymmetric distribution about the old direction, such that the new
direction makes an angle of $2 \sqrt{\DR\Delta t}$ with the old
direction \cite{Berg.1983}. The bacterial cell is considered to be
tracking the algal cell successfully for so long as it remains within
the surface $r=R$, and is considered to have lost track of the algal
cell if it strays outside this surface. Each simulation is terminated
at either a pre-determined time, or when the bacterial cell loses track
of the algal cell, whichever comes sooner.

Unless otherwise stated, the default parameter values used in the
simulations are: algal cell radius $a = 3\, {\rm \mu m}$, reversal
distance $R = 6 \,{\rm \mu m}$, algal cell swimming speed $v_{\rm alg}
= 40\, {\rm \mu m\,s}^{-1}$, bacterial cell swimming speed $v_{\rm bac}
= 200\, {\rm \mu m\,s}^{-1}$, rotational diffusivity of bacterial cell
$\DR = 0.062\,{\rm \mu m^2\,s}^{-1}$, and bacterial cell slenderness
ratio $\eta = 10$. For each set of parameters, 500 simulations are run.
We then calculate the `empirical survival function' $S(t^*)$, defined
as the fraction of simulations in which the bacterial cell is still
tracking the algal cell at dimensionless time $t^*$. The standard error
$S_E(t^*)$ of $S(t^*)$ is estimated by
\begin{equation}
   S_E(t^*) = \sqrt{S(t^*)[1-S(t^*)]/N},
\end{equation}
where N is the number of simulations \cite{Lawless.2003}.

\section{Results}

\label{sec:results}

\begin{figure}
   \centerline{
      \includegraphics[width=0.9\linewidth,height=!]{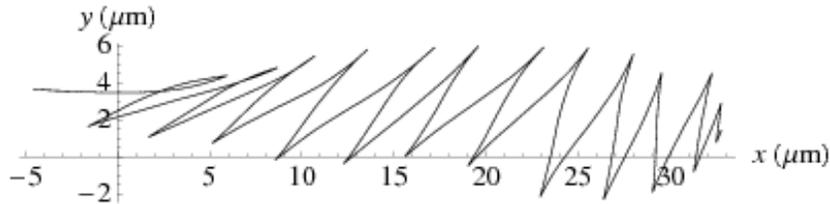}
   }
   \caption{
      Example simulation trajectory of the model bacterial cell,
      projected onto a plane. The algal cell swims along the $x$-axis.
   }
   \label{fig:example_trajectory}
\end{figure}

\begin{figure}
   \centerline{
      \includegraphics[width=0.7\linewidth,height=!]{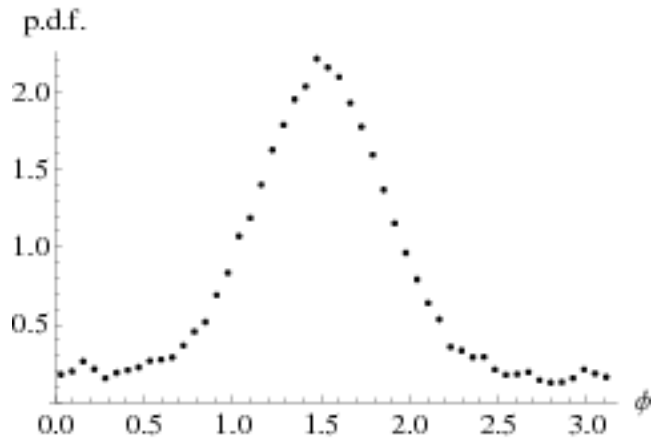}
   }
   \caption{
      Approximate probability density function (p.d.f.) for the angle
      $\phi$ between the bacterial cell's swimming direction $\bs{p}$ and the
      algal cell's swimming direction, for a set of 500 simulations
      with the default parameter values. The p.d.f is normalised so
      that in the case of isotropically distributed $\bs{p}$ it
      would be equal to unity for all $\phi$. Note the peak near $\phi
      = \pi/2$.
   }
   \label{fig:phi_probability}
\end{figure}

\begin{figure}
   \centerline{
      \includegraphics[width=0.7\linewidth,height=!]{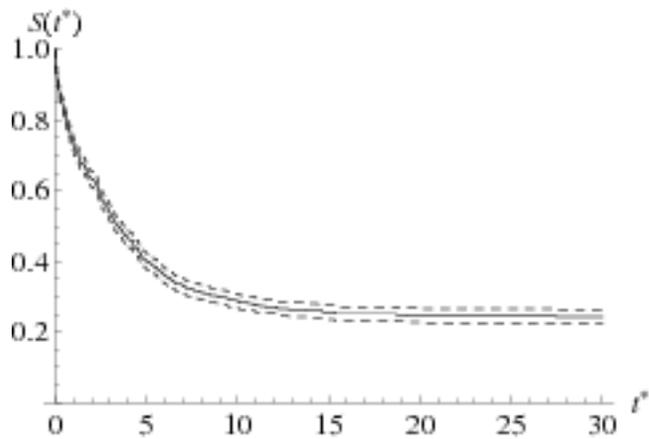}
   }
   \caption{
      Solid line shows survival function $S(t^*)$, defined as  the
      fraction of simulations in which duration of bacterial tracking
      is at least $t^*$, for a set of 500 simulations with the default
      parameter values. Dashed lines show error bounds $S(t^*) \pm
      S_E(t^*)$.
   }
   \label{fig:default_survival}
\end{figure}

\begin{figure}
   \centerline{
      \includegraphics[width=0.7\linewidth,height=!]{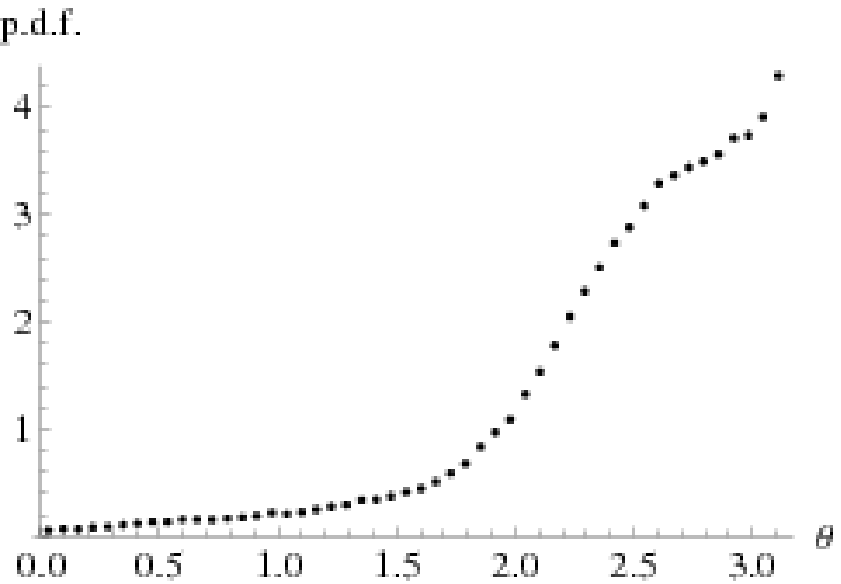}
   }
   \caption{
      Approximate probability density function (p.d.f.) for the polar
      angle $\theta$ describing the position of the bacterial cell
      relative to the algal cell, for a set of 500 simulations with the
      default parameter values. The p.d.f. is normalised so that in the
      case of an isotropic distribution it would be equal to unity
      for all $\theta$.
   }
   \label{fig:theta_probability}
\end{figure}

We shall first present data for the default set of parameter values
listed at the end of the previous section, and later look at the effect
of altering the parameter values. Figure \ref{fig:example_trajectory}
shows an example trajectory of the model bacterial cell, projected onto
a plane, for a simulation run with the default parameter values. The
trajectory bears similarities to the experimental results shown in
figure \ref{fig:fig1a_from_Barbara}, especially in that the primary
motion of the bacterial cell is almost at right angles to the motion of
the algal cell, so that the bacterial cell's motion is a zigzag. This
zigzag motion is not a once-off but occurs in most of the simulations,
as revealed by the probability density function for the angle between
the bacterial and algal swimming directions (figure
\ref{fig:phi_probability}). For the same data set, figure
\ref{fig:theta_probability} shows the probability density of a
bacterium being located at a polar angle $\theta$ in a coordinate
system centred on the algal cell, where $\theta$ is measured from the
algal cell's swimming direction. It shows that tracking most often
occurs with the bacterial cell behind the algal cell rather than in
front.

Many of the tracking durations seen in the simulations are at least as
long as those seen in experiments. Figure \ref{fig:default_survival}
shows the survival function and its standard error for 500 simulations,
using the default parameter values. By way of comparison, the \emph{P.
haloplanktis} and \emph{S. putrefaciens} bacteria in the experiments of
\cite{Barbara.2003} tracked on average for $0.7\,{\rm s}$ and
$1.8\,{\rm s}$ respectively, corresponding to dimensionless tracking
durations of $t^*\approx9$ and $t^*\approx23$. Note that the survival
function appears to reach a steady value of $S \approx 0.25$,
indicating that in a significant fraction of the simulations the
bacterial cell is able to track the algal cell indefinitely. Extending
the simulation duration up to $t^*=100$ reveals no further decay in
$S$. Closer examination of the data reveals that in all cases where the
tracking extends beyond $t^*=30$, the bacterial cell regularly collides
with the algal cell, so re-orientations due to collisions play an
important role.

\begin{figure}
   \centerline{
      \includegraphics[width=0.7\linewidth,height=!]{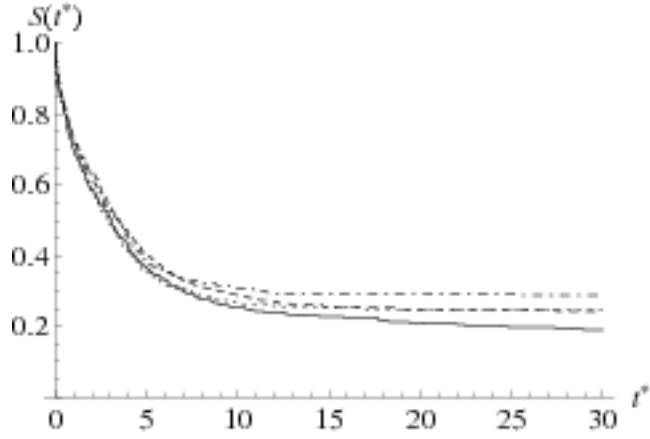}
   }
   \caption{
      Survival function $S(t^*)$ for $v_{\rm alg} = 20\, {\rm \mu
      m\,s}^{-1}$ (solid), $v_{\rm alg} = 40\, {\rm \mu m\,s}^{-1}$
      (dashed), $v_{\rm alg} = 80\, {\rm \mu m\,s}^{-1}$ (dotted), and
      $v_{\rm alg} = 200\, {\rm \mu m\,s}^{-1}$ (dot-dashed). Standard
      errors (not shown) are of the same magnitude as those in figure
      \ref{fig:default_survival}.
   }
   \label{fig:speedalg_survival}
\end{figure}

The scaling argument presented in section \ref{subsec:scaling} suggests
that tracking can occur regardless of the precise value of the algal
cell's swimming speed $v_{\rm alg}$. If the algal cell swims faster,
for instance, then its vorticity field is proportionally stronger, and
the consequent bacterial cell steering is more pronounced. In support
of this, the simulation results show that the survival function is
fairly insensitive to $v_{\rm alg}$ over a large range ($20\, {\rm \mu
m\,s}^{-1} \leq v_{\rm alg} \leq 200\, {\rm \mu m\,s}^{-1}$, figure
\ref{fig:speedalg_survival}), when tracking duration is measured by the
dimensionless time $t^*$. Since one unit of $t^*$ corresponds to the
algal cell swimming a distance equal to its own radius, the typical
distance over which tracking occurs is independent of $v_{\rm alg}$,
but the absolute duration of tracking varies inversely with $v_{\rm
alg}$. The scaling argument in section \ref{subsec:scaling} also
suggests that tracking does not depend on the precise value of the
bacterial swimming speed $v_{\rm bac}$, and indeed the survival
function was found to have no statistically significant dependence on
$v_{\rm bac}$ over the range $40\, {\rm \mu m\,s}^{-1} \leq v_{\rm bac}
\leq 400\,{\rm \mu m\,s}^{-1}$ (data not shown), while holding all
other parameters at their default values.

\begin{figure}
   \centerline{
      \includegraphics[width=0.7\linewidth,height=!]{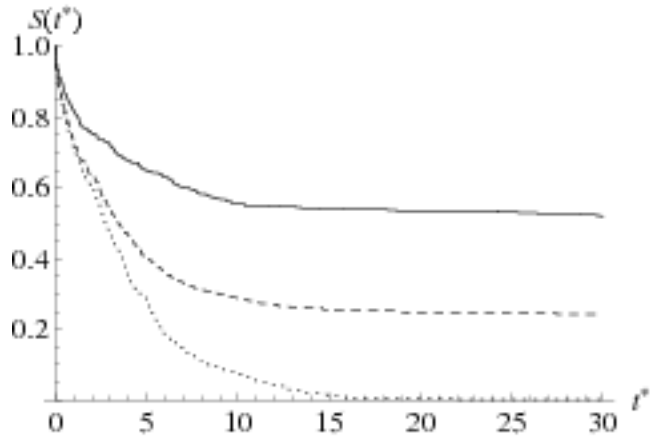}
   }
   \caption{
      Survival function $S(t^*)$ for reversal distance $R = 4\, {\rm
      \mu m}$ (solid), $R = 6\, {\rm \mu m}$ (dashed), and $R = 9\,{\rm
      \mu m}$ (dotted). Standard errors (not shown) are of the same
      magnitude as those in figure \ref{fig:default_survival}.
   }
   \label{fig:reversaldistance_survival}
\end{figure}

Where the scaling argument fails is in its prediction that tracking is
not sensitive to the reversal distance $R$. As shown in figure
\ref{fig:reversaldistance_survival}, there is a strong inverse relation
between $R$ and tracking duration. The main reason why the scaling
argument fails is because it does not account for advection of the
bacterial cell by the algal cell's velocity field, which turns out to
be important. Taking a time-average of (\ref{eq:xdot}), the
time-averaged bacterial cell velocity $\langle \dot{\bs{x}} \rangle$
has a swimming contribution $v_{\rm bac} \langle \bs{p} \rangle$ and an
advective contribution $\langle \bs{u} \rangle$, where angle brackets
denote a time-average. For simulations with the default set of
parameters, swimming contributes $\approx 30\%$ and advection
contributes $\approx 70\%$ to the time-averaged bacterial cell
velocity. For larger $R$, the advection term is smaller, and the
bacterial cell has less chance of keeping up with the algal cell.

\begin{figure}
   \centerline{
      \includegraphics[width=0.7\linewidth,height=!]{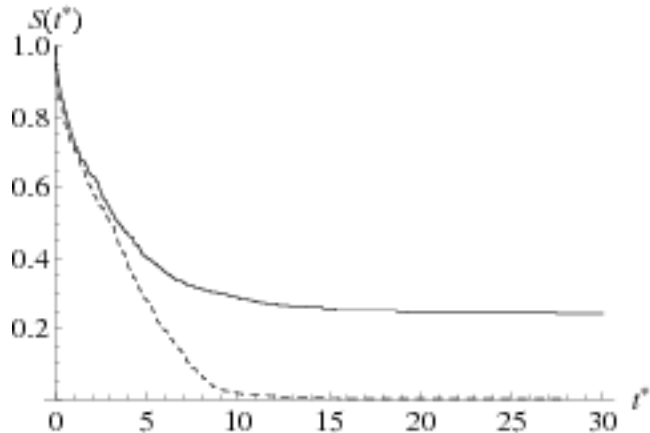}
   }
   \caption{
      Survival function $S(t^*)$ for default set of parameters (solid
      line), and for the same set of parameters but with rotation by
      vorticity and strain rate field turned off (dashed line).
      Standard errors (not shown) are of the same magnitude as those in
      figure \ref{fig:default_survival}.
   }
   \label{fig:pdot_survival}
\end{figure}

Given that advection contributes the dominant fraction of the
time-averaged bacterial cell velocity, one might ask how important
swimming and steering really are. One way to test this is to
artificially remove the deterministic terms involving $\bs{\omega}$ and
$\bs{E}$ on the right hand size of (\ref{eq:pdot}), so that there is no
`hydrodynamic steering'. The swimming direction $\bs{p}$ then evolves
only due to collisions, reversals, and Brownian rotation. As shown in
figure \ref{fig:pdot_survival}, bacterial tracking is markedly impaired
when hydrodynamic steering is turned off. Thus, even though advection
is the dominant contribution to the net bacterial motion, on its own it
is not sufficient to enable longer duration tracking. This concurs with
the observations of \cite{Barbara.2003}, who, using latex beads as
passive substitutes for bacteria, found that there was no tracking or
entrainment of beads by any algae.

\begin{figure}
   \centerline{
      \includegraphics[width=0.7\linewidth,height=!]{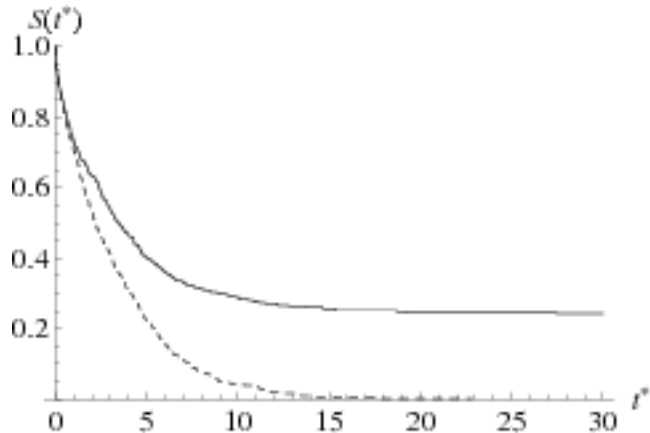}
   }
   \caption{
      Survival function $S(t^*)$ for an elongated bacterial cell
      ($\eta=10$, solid line), and for an spherical bacterial cell
      ($\eta = 1$, dashed line). Standard errors (not shown) are of the
      same magnitude as those in figure \ref{fig:default_survival}.
   }
   \label{fig:eta_survival}
\end{figure}

The rate of change of bacterial swimming direction induced by the algal
cell's strain-rate field $\bs{E}$ depends on the elongation $\eta$ of
the bacterial cell, through (\ref{eq:pdot}) and (\ref{eq:alpha}). A
spherical cell is not rotated at all by the strain-rate field, and an
elongated cell is rotated so as to align with the direction of maximum
stretching. Since the direction of maximum stretching points
approximately toward or away from the algal cell (figure
\ref{fig:strainratefield}), one expects a more elongated bacterial cell
to have a superior tracking performance. Figure \ref{fig:eta_survival}
shows that this is indeed the case. For large $\eta$, the precise value
is not important; survival curves for different $\eta$ beyond $\eta
\approx 5$ lie almost on top of one another (data not shown).

\begin{figure}
   \centerline{
      \includegraphics[width=0.7\linewidth,height=!]{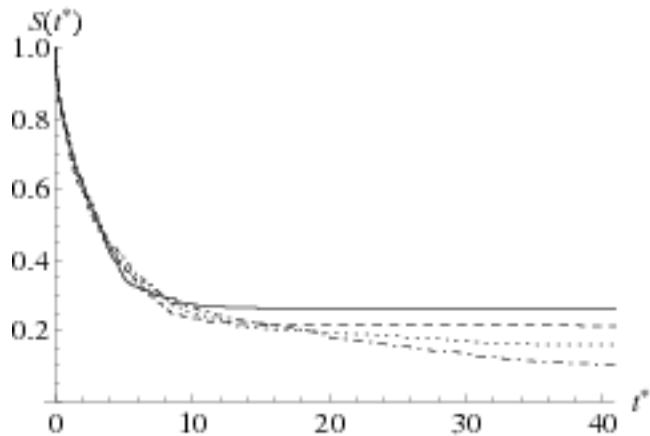}
   }
   \caption{
      Survival function $S(t^*)$ for bacteria with rotational diffusion
      coefficients $\DR$ of $0\;{\rm radians}^2{\rm s}^{-1}$ (solid),
      $0.062\;{\rm radians}^2{\rm s}^{-1}$ (dashed), $0.12\;{\rm
      radians}^2{\rm s}^{-1}$ (dotted), and $0.2\;{\rm radians}^2{\rm
      s}^{-1}$ (dot-dashed). Standard errors (not shown) are of the
      same magnitude as those in figure \ref{fig:default_survival}.
   }
   \label{fig:dr_survival}
\end{figure}

As explained in section \ref{subsec:modelbac} the rotational diffusion
coefficient $\DR$ for the bacterial cell is difficult to estimate,
because it may be due to `wobbly swimming' rather than true, thermal,
Brownian rotation. Figure \ref{fig:dr_survival} shows that in any case
there is only a weak dependence on $\DR$. A large value of $\DR$ does
not significantly affect the survival function at early times $t^* <
20$, but causes a slow decay in the survival function at longer times,
so a bacterial cell cannot continue tracking indefinitely.

\section{Discussion}
\label{sec:discussion}

The simulations presented in this paper suggest that the bacterial cell
steering seen by Barbara and Mitchell \cite{Barbara.2003} could be a
passive hydrodynamic effect. The model bacterial cell in the present
study does not actively steer, it simply swims forward and back, and
through its interaction with the model algal cell's velocity, vorticity
and strain rate fields (all three of which are important) it is able to
track the model algal cell. The simulated bacterial cell trajectories
have a similar zigzag shape to those seen in experiments, and some of
the tracking durations seen in the simulations are at least as long as
those seen in the experiments. Indeed some tracking events last
indefinitely (for as long as the simulations are run), though in
reality such long tracking events would be disrupted by effects such as
unsteadiness in the algal cell's velocity field, changes in the algal
cell's swimming direction, or noise in the bacterial cell's
chemosensory system.

The simulations indicate that the reversal distance $R$ (the distance
from the algal cell at which the bacterial cell reverses direction) has
a very strong effect on tracking duration, with smaller $R$ enabling
longer tracking duration. The reversal distance is an artificial
construct of the model, since in reality the bacterial cell's
chemotactic response is almost certainly more complicated than simply
reversing direction whenever the chemoattractant concentration drops
below a certain threshold. Even so, the inverse relation between $R$
and tracking duration shows that the ability to respond rapidly to
chemoattractant signals and reverse direction with high frequency is
crucial in order for tracking to occur, and this may represent an
evolutionary pressure on certain marine bacteria toward faster response
times.

Another result of our simulations is that the dimensionless tracking
duration (proportional to the number of body lengths the algal cell
swims while being tracked) is largely independent of the algal and
bacterial swimming speeds. This seems at odds with the experimental
results, where \cite{Barbara.2003} found that bacteria increase their
swimming speed by roughly a factor of two when tracking (relative to
not-tracking). If tracking occurs just as well at lower bacterial
swimming speeds, why do the bacteria swim faster? First, the faster
swimming speed increases diffusive flux, and hence nutrient
acquisition, by about $30\%$ \cite{Barbara.2003}. Second, it could also
be that faster swimming speeds result in more robust tracking when the
algal cell changes swimming direction, a complication not included in
our model. Third, it could be that some bacteria in the culture were
committed to anaerobic metabolism, and only the faster-swimming
aerobically metabolising cells performed tracking (J. G. Mitchell,
personal communication).

Perhaps the most serious limitation of the present model is that it
treats the bacterial cell as being infinitesimally small, whereas in
reality its length is comparable to the algal cell diameter. Thus,
instead of being rotated by the vorticity and strain-rate fields at a
point in space, in reality the bacterial cell `samples' these fields
over its length. In defense of the present model, we point to the work
of \cite{Visser.2000}, who performed a combined experimental and
theoretical investigation of the re-orientation of elongated food
particles (diatoms) in copepod feeding currents. The food particles in
their experiments were of similar length to the copepods generating the
currents, but the authors still found convincing agreement with
theoretical calculations that treated the food particles as
infinitesimal.

Given the importance of the algal cell's vorticity and strain rate
fields in re-orienting the bacterial cell, it is natural to ask whether
ambient, turbulent fields might disrupt tracking. However, the strength
of the strain rate tensor due to turbulence varies from about
$1.5\,{\rm s}^{-1}$ in the upper mixed layer of the ocean under strong
wind forcing to $0.005\,{\rm s}^{-1}$ at the thermocline
\cite{Luchsinger.1999}, and these magnitudes are small compared with
the strain rate of $v_{\rm alg} / R \approx 7 \,{\rm s}^{-1}$ in the
vicinity of the algal cell. Thus, turbulence is unlikely to disrupt
tracking. It might, of course, play a role in how the bacterial cell
finds the algal cell in the first place, which is not addressed in our
model. Another consideration related to the strength of the algal
cell's strain rate field is whether it might be strong enough to deform
the bacterial cell's flagellum. An order of magnitude estimate using
the bending stiffness reported by \cite{Darnton.2007} for an \emph{E.
coli} flagellum suggests that deformation could be significant. This
provides another potential mechanism for passive hydrodynamic steering,
which has not yet been explored.

The idea that microorganisms may exploit vorticity and strain rate
fields in their feeding behaviours is not new. In \cite{Pedley.1987},
it is suggested that certain freshwater Crustacea (cladocerans), which
feed by intermittently swimming vertically upward and sinking down
again, might enhance their prey capture rate through hydrodynamic
effects. Their calculations show that bottom heavy motile algae are
focussed into the wake of one of these sinking crustaceans by its
vorticity field (`gyrotaxis'), whereupon the crustacean presumably
consumes the algae when it swims up. In \cite{Timm.1995}, a similar
analysis is performed for the case of a bottom heavy flagellate
focussed into the wake of a nonmotile, sinking algal cell. In the
simulations of \cite{Luchsinger.1999}, elongated model bacteria using a
run-and-reverse (`back-and-forth') motion are able to stay close to a
nutrient source even at high ambient shear, because the strain rate
field aligns the bacteria toward the nutrient patch.

While the simulations presented in this paper suggest that the apparent
steering seen by \cite{Barbara.2003} could be a passive hydrodynamic
effect, the results do not rule out the possibility that the bacteria
are actively steering as well. The fact that bacteria were able to
follow `ghost tracks' some distance behind algal cells (mentioned at
the bottom of p82 of \cite{Barbara.2003}) suggests that there may also
be active steering, since vorticity and strain rate fields far behind
the cell are probably not sufficient to rotate the cell. If \emph{P.
haloplanktis} and \emph{S. putrefaciens} are indeed capable of
steering, then they are not the first bacteria to be found to do so.
Experiments indicate that another (unnamed) species of bacteria can
sense an oxygen gradient over its body length and steer relative to the
gradient in a continuous fashion \cite{Thar.2003}, though it is quite
different in shape from \emph{P. haloplanktis} and \emph{S.
putrefaciens}. In another example, \emph{Rhodobacter sphaeroides} has a
similar shape to \emph{P. haloplanktis} and \emph{S. putrefaciens},
with a single flagellum, and is able to changes direction by altering
the conformation of this helix \cite{Armitage.1999}, though there is no
evidence of directed steering in this case.

Finally, we should note that it has not been our intention in this
report to make light of the work of \cite{Barbara.2003}. Their finding
of tracking bacteria is interesting in its own right and nutrient
turnover by tracking bacteria could play a key role in the food web
especially if, as the authors suggest, a similar mechanism applies to
the tracking of marine snow. Rather, our intention has been to
elucidate the intriguing possibility that not only are the marine
bacteria able to perform tracking, but that they do so by exploiting
the velocity, vorticity, and strain-rate fields of their quarry.

\section*{Acknowledgements}
J. T. Locsei is supported by an Oliver Gatty Studentship from the
University of Cambridge. The authors are grateful to J. G. Mitchell for
helpful discussions.

\bibliographystyle{plain}
\bibliography{everything}

\end{document}